# A Survey on Formal Verification Approaches for Dependable Systems

### Fayhaa Hameedi Khlaif
PG Scholar, College of Engineering, Department of Computer Engineering, Mosul University, Iraq
Email: fayhaa.20enp14@student.uomosul.edu.iq

### Shawkat Sabah Khairullah
Lecturer, College of Engineering, Department of Computer Engineering, Mosul University, Iraq
Email: shawkat.sabah@uomosul.edu.iq

**Abstract:** *The complexity of digital embedded systems has been increasing in different safety-critical applications such as industrial automation, process control, transportation, and medical digital devices. The correct operation of these systems relies too heavily on the behavior of the embedded digital device. As a result, any mistake or error made during the design stage of the embedded device can change the overall functionality of the critical system and cause catastrophic consequences. To detect these errors and eliminate their effects on the system, new error detection approaches must be innovated and used in the design of the digital system. However, these methods require enormous costs and time. One of these methods being employed to solve this issue is called Verification and Validation (V&V) which confirms that the system behavior meets the requirements early in the development process, before moving on to the implementation phase. Because of their benefits and importance in the building of complex digital systems, the employment of formal V&V methods has recently attracted a lot of attention. This paper focuses on presenting various studies on formal verification approaches and how the V&V can be achieved for developing high dependable digital embedded systems.*

**Keyword:** *safety-critical, embedded device, error, formal verification, verification and validation.*

## 1. INTRODUCTION

A digital device is a piece of cyber-physical equipment that uses digital data or an electronic device that can create, generate, send, or process information. The core of the digital devices is Embedded system, which is a microprocessor-based computer hardware system with software that is intended to execute a specific task, either independently or as part of a larger system. Embedded system applications range from simple digital devices such as watches and microwaves to digital devices in large and complex systems such as hybrid vehicles and avionics.

The systems that combine cybernetic aspects with physical processes are described as cyber-physical systems (CPSs). Physical processes are monitored and controlled by embedded computers, which usually have feedback loops, and physical processes impact computations. Sensors in the system allow it to sense the physical environment around it and any event sensing by it will react to it using control and actuator modules in response to a physical change in the system. As a result, a CPS is made up of a collection of modules that work together. Those modules normally run concurrently, allowing numerous operations to be performed at the same time [1].

Different applications and systems are used CPS such as driverless vehicles, biomedical systems, healthcare systems, smart grids, and a variety of industrial applications. Systems and control engineers have made great progress in developing methodologies and tools for system science and engineering throughout the years (e.g., time and frequency domain methods, state space analysis, filtering, prediction, optimization, robust control, and stochastic control). Simultaneously, researchers in computer science and software engineering have made significant advances in program verification and validation. Validation ensures that a system fits the needs of clients, whereas verification determines whether a system meets its specification.

The complexity of representing both the cyber (e.g., software, network, and computing hardware) and the physical (e.g., the environment) aspects of any CPS makes it even more difficult to model (physical processes and their interactions). If we don't expect the physical and logical components to fail in tandem, simplified models are easily invalidated[2]. As a result, it is necessary to verify the specification before moving on to the next stage of system development, especially when developing a reliable and safe system. As resulted, the errors will be discovered early, therefore the speed of correction will be faster as well as reducing costs. To achieve that in the early stage of system development, formal methods are applied, whereby using model checking technique, specifications can be checked against some user-defined requirements given as temporal logical formulas. This answers the question of whether the requirements are fulfilled, and if they aren't, appropriate counterexamples are generated to hunt down the undesirable state[3].

The remainder of this paper is organized as follows: Section 2 introduced the concept CPS and its basic building modules. Also, this section discussed the verification and validation fundamentals. Modeling systems and modeling languages are illustrated in section 3, while section 4 discusses the formal verification and focuses on explaining the model checking technique. Section 5 demonstrates fault classification, categories of the fault, and normal testing, as well as the critical role of fault tolerance in dependability systems. Section 6 illustrates the difference between normal testing and formal verification. The differences between runtime and offtime verification are discussed in section 7. In section 8, a review of the literature on the verification and validation process is presented. Section 9 is devoted to the conclusion of the paper.

## 2. BASICS
### 2.1 How Can Scientific Computing Achieve Credibility?
The term "credibility of computational results" refers to the ability to believe or trust the results of an analysis. The following are the basic aspects that must be considered in order to ensure the credibility of computational results:

- The work quality of the analysts
- The physics modeling quality
- The verification and validation processes [4]
- The qualification level of analyses and equipment [2]. And
- Sensitivity analyses and uncertainty quantification [1]

### 2.2 Verification
Is the process of evaluating the products of a software development phase to provide assurance that they meet the requirements defined for them by the previous phase.[4]

### 2.3 Validation
Is the process of testing a computer program and evaluating the results to ensure compliance with specific requirements.[4]

### 2.4 V&V Process
The Verification and Validation process is a technical process of systems, software, and hardware engineering. The relationship between these processes is interconnected and can be expressed together as (V&V). When the organization applied V&V processes on its system, this will be assisted it to build quality into the system and obtain an objective evaluation of products and processes throughout the life cycle. This evaluation shows whether the specifications are correct, complete, accurate, testable, and consistent. The V&V processes determine whether the development products of a given activity conform to the requirements of that activity and whether the product satisfies its intended use and user needs. The determination includes the assessment, analysis, evaluation, review, inspection, and testing of products and processes. V&V is performed in parallel with all life cycle stages, not at their conclusion[5].

## 3. SYSTEM MODELING
System modeling is the initial, and often crucial, step-in verification. The right choice of model and modeling language is important for both designers and users of verification tools. Modeling Languages, it is useful to distinguish between formalisms, which are mathematical objects, and concrete modeling languages (and tools) which support such formalisms. A plethora of modeling languages exist, developed for different purposes, including:

- Hardware description languages (HDLs).
- General-purpose modeling languages, such as UML and SysML.
- Architecture Description Languages (ADLs), such as AADL.
- Simulation-oriented languages and tools, such as Matlab-Simulink or Modelica. Simulink and related tools provide primarily simulation, but also code generation and even formal verification in some cases.
- Reactive programming languages, such as the synchronous languages Lustre and Esterel. These languages have been developed specifically for formal verification purposes, using model-checking or theorem-proving techniques, including satisfiability solving[6].

## 4. FORMAL VERIFICATION
### 4.1 Formal Verification (FV)
Is the analysis of the behaviors of a design using mathematical analysis tools instead of calculating results for specific values. This means that rather than testing specific values, an FV tool will examine at the whole space of possible simulations. Of course, it won't execute all potential simulations, but it will utilize advanced mathematical approaches to consider all their potential behaviors. The distinction between simulation and FV is that simulation examines individual points in the space of possible tests, whereas FV examines the entire space at once. You can probably already see the incredible power we can get by employing this strategy from this simple definition[7]. Formal verification is a systematic approach for validating that a design implementation meets the requirements of its specification using exhaustive algorithmic techniques [8]. Figure 1 illustrate the formal verification process.

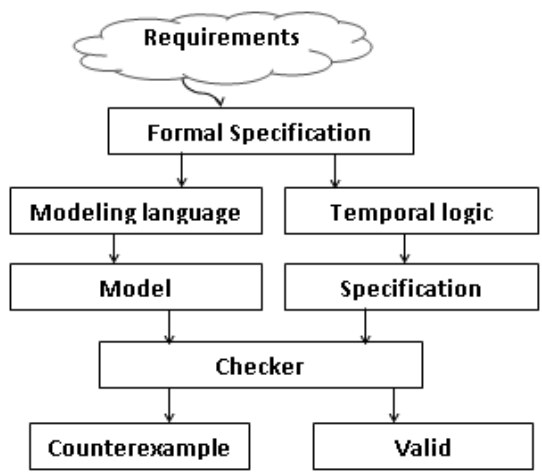

*Figure 1 Formal verification process*

### 4.2  Formal Verification Methods

Methods of formal verification can be accomplished in a variety of techniques, each with its own set of advantages and disadvantages. Theorem proving and model checking is the most commonly used in the field of smart contracts. [8]

### 4.3 Model Checking

Is a computer-assisted method for the analysis of dynamical systems that can be modeled by state-transition systems. Drawing from research traditions in mathematical logic, programming languages, hardware design, and theoretical computer science, model checking is now widely used for the verification of hardware and software in industry. In addition, Uses of temporal logic and data-flow-analysis techniques have also made model checking more naturally efficient[6].

Every embedded system must be designed to meet certain requirements. Such system requirements are also called properties or specifications. A mathematical specification of system properties is also known as a formal specification. The specific formalism will be used is called temporal logic. As the name suggests, temporal logic is a precise mathematical notation with associated rules for representing and reasoning about timing-related properties of systems. While temporal logic has been used by philosophers and logicians since the times of Aristotle, it is only in the last thirty years that it has found application as a mathematical notation for specifying system requirements. One of the most common kinds of system property is an invariant. It is also one of the simplest forms of a temporal logic property[9].

An invariant is a property that holds for a system if it remains true at all times during operation of the system. Put another way, an invariant holds for a system if it is true in the initial state of the system, and it remains true as the system evolves, after every reaction, in every state. In practice, many properties are invariants. [9]

System properties may be safety or liveness, a safety property guarantees that something wrong will never happen, whereas a liveness property ensures that something good will eventually happen[10].

### 4.4  FORMAL MODELING OF FAULTS

Reliability is one of the most important characteristics of the system quality. It is defined as the probability of failure-free operation of the system for a specified period of time in a specified environment. The system reliability has become an important design aspect for computer-based system, and it is considered of primary importance in systems deployed in safety critical applications such as avionics, military, aerospace and transportation. In such systems, the cost of the unreliability is very high, a fault can damage the whole system and leads to catastrophic failures. There is a very large set of different failure  sources for the system components[11].

### 4.5  Fault Classification

In everyday language, the terms fault, failure, and error are used interchangeably. In fault-tolerant computing parlance, however, they have distinctive meanings. A fault (or failure) can be either a hardware defect or a software/programming mistake (bug). In contrast, an error is a manifestation of the fault/failure/bug. Both faults and errors can spread through the system[12].In general, a failure represents the condition in which the system deviates from fulfilling its intended functionality or the expected behavior [33]. A failure happens due to an error, that is, due to reaching an invalid system state. The hypothesized cause for an error is a fault, which represents a fundamental impairment in the system. The notion of faults, errors, and failures can be represented using the following chain [34].

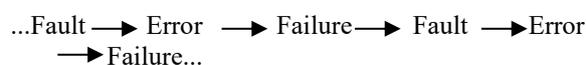

Each component of the system is susceptible to be affected by internal faults (fault occurring during the system design, the system manufacturing or due to aging-effect) or external faults (faults due to environmen-tal perturbations). The fault can occur either at the software layer or the hardware layer of the system. Therefore, we differentiate two categories of faults.

### 4.6 Fault Categories

The hardware faults are caused by physical phenomena affecting the hardware components, such as environmental perturbations, manufacturing defects, and aging-related phenomena.

The software faults can only impact the software components. However, the hardware faults can propagate through the different system layers, and affect both the hardware and the software[11].

### 4.7 Fault Tolerance

Fault tolerance is defined as the ability of the system to perform its function even in the presence of failures [34][36]. This implies that it is utmost important to clearly understand and define what constitutes the correct system behavior so that specifications on its failure characteristics can be provided and consequently a fault tolerant system can be developed[13].

### 5. NORMAL TESTING VERSUS FORMAL VERIFICATION

Testing is a dynamic technique that involves running a (software) system under specific conditions (predefined environment/input sequences) and determining whether the observed behavior differs from the required behavior. The goal of model-based testing is to check that the implementation's behavior confirms that of the specification model through execution[14]. In model-based testing (MBT), the aim is to check by execution that the behavior of implementation conforms to that prescribed by the specification model. The primary idea behind MBT is that instead of manually building test cases, a chosen algorithm generates them from a model automatically. MBT is most commonly associated with the automation of black-box test design, but it has lately been used to white-box tests as well [15]. Formal verification differs from normal testing in the following ways:

It occurs during the design phase, whereas testing occurs after the code has been created.

It guarantees that there are no errors in the intended design after implementing it.

Testing for every single input, a chain of inputs, unanticipated events, is almost too difficult [8]. To demonstrate this point, consider testing the functionality of a simple 16-bit comparator, as illustrated in Figure 2. the specifications of this example is very simple:

Y1 is always a logical 1 whenever X1 > X2.
Y2 is always a logical 1 whenever X1 ==X2.
Y3 is always a logical 1 whenever X1 < X2.

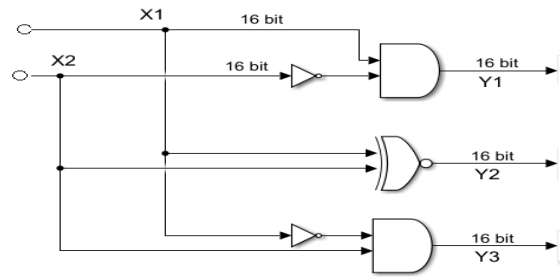

*Figure 2 16-bit comparator*

232 vectors are required to thoroughly test our comparator example via simulation. Assume that our simulator can evaluate a single vector per microsecond. To finish this verification procedure using normal testing, it would take 1.2 hour, what if the inputs are 32 bits or 64 bits, or if there are a lot of inputs? but when applying Formal verification technique, the vector set will be reduced because of removing redundancy and focusing on selecting targeted.

### 6. RUN TIME VERIFICATION VS OFFTIME VERIFICATION

Runtime Verification (RV) has been referred to a variety of names, including runtime monitoring, trace analysis, dynamic analysis, and so on. The term "verification" indicates that something is correct in terms of some property. This differs from the term monitoring, which simply implies that some type of behavior is being observed. Some people consider monitoring to be more particular than verification because it implies interaction with the system, whereas verification is a passive process. RV is a simple, yet rigorous, formal method that complements traditional exhaustive verification techniques (such as model checking and theorem proving) with a more practical approach that analyzes a single system execution trace. RV can provide very detailed information on the runtime behavior of the monitored system at the price of restricted execution coverage[16]. The general scheme [17] is given in Fig. 3. The property can be expressed as a set of rules, a program, or a formal specification language (automata, logic formula, grammar) [17].

A runtime verification framework's monitor is a key component. A component executed along with the system for the purposes of the runtime verification process is referred to as a monitor[18].

A monitor is made up of a set of properties. The monitor's job is to observe the system's operation and generate a verdict, or a statement if the observation meets the properties. A series of system states or a series of input and output events could be observed. Monitoring is done either step by step while the system is running or over a log that records the observations[17].

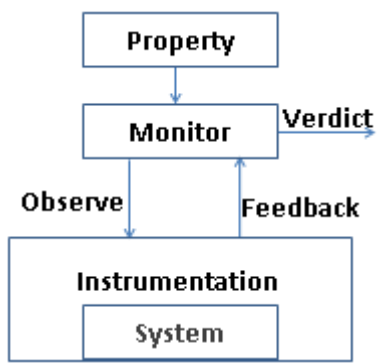

*Figure 3 General scheme of runtime verification*[17]

Instrumentation is the process of probing and extracting signals, traces of events, and other relevant information from a software or hardware system though it is executing. Instrumentation is a crucial part of a runtime verification setup since it allows monitors to be hooked to the system.

The type of system to be monitored influences the instrumentation techniques used. For example, monitoring hardware systems may include probing mixed-analog signals with physical cables, but software instrumentation is strictly related to the programming language used to implement the software or the low-level language used to compile it [16].

## 7. LITERATURE REVIEW

In Table I we can see a summary of approaches that have been used for verifying system properties. Askari Hemmat et al. (2015) [19], proposed an approach to verify and implement complex system design. The abstraction models must be used to verify functionality in order to decrease the number of potential bugs in the implementation stage. Then one of the formal verification methods should be utilized to ensure that the design of a complex system is implemented with limited risks, such as the model checking technique, which determines whether the design meets the desired properties automatically.

They utilized a train control system as an example for a complex system and showed how to model the system. [19] determines the limits of the safe speed and acceleration in the modeling system and they used the NuSMV technique to verify the correctness of the model properties. After that, the verified model was implemented on an ARM Cortex processor.

The system is controlled automatically as well as on-board drivers have a limited role to play during normal operation. The trains are controlled in their velocity by a station computer that is part of the Advance Automatic Train Control (AATC) system, which uses an algorithm to adjust the speed and acceleration of each train. [19] described an algorithm for modeling train operation that is based on another algorithm that is employed by the station computer and is responsible for generating the appropriate train velocity and acceleration commands. This algorithm specified the safety property. Based on this algorithm, [19] modeled the system in AF3 (AutoFocus3) to verify system operation before generating code. The AF3-modeled system includes a monitor block that represents the position, velocity, and acceleration of the trains.

This block contains the system's simulation and verification processes. In simulation, the train's initial position is calculated in a way that avoids violating safety regulations. In verification, the property is verified if the train's initial position is greater than the Worst Case Scenario Distance (WCSD) to the heading obstacle. Finally, the train control algorithm was implemented on an ARM CORTEX -M4 platform after the verification model confirmed that the system specification was not violated. They concluded that their method can be used to verify and implement another system of high complexity.

Also, CPS is considered complex system and needs to use formal verification method that achieve verification and validation objectives. Kang et al. (2017) [20], have presented an approach to guarantee that CPS operates in safety and effectively in the automotive domain. They used EAST-ADL language for architectural specification and timing behavior constraints for automotive embedded systems. Formal analysis on both functional- and non-functional properties, is performed using Simulink Design Verifier (SDV)and UPPAAL-SMC (formal verification tool). They demonstrate their approach by using the autonomous traffic sign recognition vehicle as case study.

In SDV, each property is modeled as a Simulink block or embedded MATLAB function (EML) which can be valid or invalid. After analyzing the counter-examples that generated by Both SDV and UPPAAL-SMC, Kang et al. (2017) [20], have discovered the reason of errors. Therefore, the invalid property and system model have been refined based on the error traces. After the change, the invalid property became valid.

Due to the lack of stochastic modeling notations and descriptive blocks required during proof creation, SDV provides limited support for modeling and analysis of XTC and energy constraints. For the verification of those restrictions, they have been used UPPAAL-SMC, and then the results were validated in 95% confidence.

CPSs handle discrete and continuous signals, so modeling the behavior of such systems needs a method that can cope with both: continuous and discrete systems. As a result, hybrid models are required to capture both discrete and dynamic characteristics in Cyber Physical Systems. Kumar and Kumar. (2016) [21], used PROMELA which is used to model timed automata and verify them using the Spin model checker. Timed automata can be used to effectively illustrate the behavior of CPS protocols. SPIN comes very handy for protocols that operate in lossy settings. SPIN model checker in its original form cannot simulate passage of time, which is crucial for modeling timed automata. As a result, [21] proposes a validation model for timed automaton-based systems that uses the Spin model checker and applies it to a timed automaton-based traffic controller. A timed automaton is a hybrid system that combines a clock that keeps track of the passage of time with a discrete pure signal. Kumar and Kumar attempted to model this system using PROMELA's basic constructs and attempted to verify the safety and validity properties of a traffic controller modeled as a hybrid system. Using Spin Model Checker, this work gives an easy and effective approach to build protocols for timed automaton-based systems. Eventually, this work will help protocol designers and software developers to model and verify bug-free protocols for Timed Automata-based Hybrid Systems.

It is also very important to assure that the safety-critical system (SCS) is working correctly. The SCS requires fault tolerance [22], which ensures that the system continues to function properly even if one or more components fail. Redundancy can provide tolerance for a single failure. If one of the redundant subsystems fails, the remaining redundant subsystems will continue to perform the required functions.

Pakonen and Buzhinsky. (2019) [22], provided a method for employing model checking to verify the fault tolerance for instrument and controls (I&C). As an example for a SCS, [22]utilizes the reactor protection system of the projected US.EPR nuclear power plant. The protection system for a fault-tolerant I&C system is arranged into four redundant independent divisions and is located in separate buildings.

They have been modeled the function block diagram design by using the graphical editor of the MODCHK tool, and by the text editor of this tool, they could specify the properties. As well, this tool has another advantage that is producing the necessary input files for NuSMV and visualizing the counterexample generated by NuSMV with an animation view of the function block diagram. The formal property of (un)desired behavior is stated using temporal logic languages such as LTL, CTL, and PSL, and the system model is expressed as a finite states machine.

The verification procedure in safety systems is critical, so [22] focused on verifying single fault tolerance in open-loop models rather than describing a full failure model. The failure model is implemented in NuSMV by adding modules to all signals on one division. They've made a list of several functional requirements, some of them have been formulated using CTL, which have been specified as non-deadlock requirements, and the remaining requirements have been formulated using LTL/PCL and CTL. In case of exception delays, the requirements have been formulated in LTL/PCL. Because of a huge number of states in the I&C system, [22] needed to use Bounded model checking (BMC) which is a feature of NuSMV that limits the length of the checked state transition sequences, and they focused on avoiding the explosion state space, which is the primary challenge in the case example. After model checking runs, the results demonstrate that the cases without delays were best handled by BDD – based model checking algorithm than BMC and showed that the absence of deadlock requirements can be proven via CTL model checking, whereas only BMC was able to check the requirements in the case of delays. As a result, in the case of delays, checking for the absence of deadlock requirements was impossible.

Dependability analysis such as safety, reliability, efficiency, etc. is necessary for evaluating the design of the safety-critical system to ensure that it was modeled correctly and verified before being deployed in a real-time scenario. Therefore, it needs to model it using Various modeling techniques such as unified modeling language (UML), fault tree, failure mode effect analysis, and reliability block diagrams (RBDs), which are suitable for modeling all static properties, while they would fail when used to model dynamic properties. The extension of RBD which is named as dynamic reliability block diagram DRBD can be used for modeling the dynamic behaviour system. Kumar et al. in 2021[23], presented an approach for model-based verification using DRBD. To analyze, check and verify the features of safety and reliability such as nonliveness, deadlock, faulty state, or error design that cannot be done manually using DRBM modeling, therefore they used color Petri net CPT for complete proof design. They utilized a digital feedwater control system (DFWCS) of a nuclear power plant (NPP) as case study. The DRBD for the DFWDS has been illustrated in detail in [23]. To analyze and verify the design of the DRBD model for DFWCS is correct, the components in DRBD must be represented in CPT as states which can take one of the following forms: active, failed, or standby. The CPT model representation from the DRBD model has been illustrated in detail in [23]. After creating the CPT model for DFWCS, the simulation step has been performed using the CPT tool which provides a detailed report for the model system. This report introduces information about the state space for the CPT model in terms of statistics and liveness properties. As for liveness, it demonstrated that there are three dead marking states and two dead transitions. By this analysis that verified the CPT model, the design errors in the DFWCS have been discovered and this mean that there are errors in the DRBD model. To remove the deadlocks that have been stated above, the DRBD need to modify it then modified the CPT model. After simulating the modified DFWCS model, the result showed that there is no found any deadlock whether marking or transition instances this ensures the correctness of the modified CPT. Eventually, they found that the designed model is effectiveness model with full proof.

It is preferable to employ formal methods to verify the model design for SCS and prove its complete cor-rectness, which is lead to improving SCS's dependability qualities (safety and reliability are the more important dependability parameters). Therefore [24] applied the NuSMV model checking technique on Mode Transition Logic MTL of an autopilot system to analyze the verification status of the model. NuSMV tool has been used for verifying the vertical mode functionality of the MTL (represents a very critical functionality in aircraft).

The properties of MTL that are required to verify it informal method, need to specify it using temporal logic language such as Computation Tree Logic (CTL) or Linear Temporal Logic (LTL) which are used to check the model for properties. After carrying out the model State-space and getting a detailed analysis of the model in terms of numerous parameters from the test cases, the specifications are simultaneously provided to NuSMV in the form of an input code with the smv extension. The program's truth or falsity is determined by compiling and simulating the code. Then the results of the checked property which have been got by NuSMV, are compared to those received by Simulink. If any corrections are found, the model is re-verified and analyzed. Otherwise, the model's efficiency is determined.

EAST-ADL is targeted to the design of automotive embedded systems, with a particular emphasis on structural and functional modeling[25]. The behavior is defined in terms of functional blocks at the EAST-ADL component abstraction level and, because a component's functional activity is represented using external notations like Simulink or UML, the capacity to create, verify, and transform EAST-ADL models using formal methods is limited. Furthermore, because many automotive functions are real-time, formal verification of both functional and timely behavior is required at the architectural level to verify that real-time requirements have been met. As a result of the behavioral description not being integrated into the execution semantics, EAST-ADL models are more difficult to transform, analyze, and verify models. Therefore, [25] presents a way for integrating architectural description languages and verification methodologies (based on Eclipse plug-ins) by implementing it in the tool ViTAL, which is suited for EAST-ADL models. ViTAL is improving the EAST-ADL language's behavioral definition, providing for formal modeling, simulation, and verification of the requirements (functional and timing). The EAST-ADL architectural model and the created TA behavioral model are required artifacts for the formal analysis of the system. In [25], the information regarding model transformation to UPPAAL port has been covered in detail. Enoiu et al. - 2012 have been integrated such models into ViTAL, so that they can simulate and verify whether a certain requirement is satisfied, by model verifying the TA description using UPPAAL PORT.

They conducted a case study in a Brake-By-Wire (BBW) system which was modeled in EAST-ADL to demonstrate the usage of ViTAL.

In the tool ViTAL, they modeled, simulated, and verified the BBW system. Papyrus UML Editor was used to model the system, and a UML profile was created to describe the architecture. The safety and liveness properties of the BBW system, have been verified and the property of deadlock freedom is satisfied for all execution paths of the state-space. They used the CTL specification as an example of an architectural property because the tool ViTAL provides simulation and verification of architectural properties. The CTL specification ensures the brake reaction delay stated in the BBW model. They point out that during verification, ViTAL is indirectly utilizing the delay constraints information from EAST-ADL models so they found that this delay constraint information should be considered a transformation parameter and then verified automatically in UPPAAL PORT to handle automated integration during verification of TCTL (Timed Computation Tree Logic) properties.

During the system's design cycle, the verification process can be conducted with different stages. Wisniewski et al. (2019) [26], verified the system's operation in three stages before the final implementation in hardware. The stages have been represented as follows: specification stage, modeling stage (which represents software verification), and finally hardware verification that conducted after the implementation process in a programmable device. Khan et al. applied their approach to a cyber-physical system (CPS) and illustrated it using a direct matrix converter (MC) with transistor communication and space vector modulation (SVM). In this case study as mentioned hereinabove, CPS is verified through a series of stages.

Initially, the (SVM)algorithm was specified using a Petri-net, which can be live and safe. This stage involves creating a Petri-net based on the requirements. As a result, there are 49 places and 31 transitions in the net. To ensure that all places of the net are reachable, and the Petri net is live, the model checking technique should be applied in the next phase. In this phase the first step is the simple rule-based logical model (RBLM) which is utilized to formally verify the net. To perform formal verification, the net is first written on (RBLM), where the variables are Petri-net places and all the Petri-net transitions are written as separate rules, each one contains a precondition and postcondition. Following the completion of the RBLM, the developed m2vs tool was utilized to generate an automatically verifiable model in NUXMV format, as well as based on the RBLM, a synthesizable prototype model, is generated in the VHDL language. After formal verification, the initial Petri-net specification is analyzed such as liveness, safety, and the number of reachable states. When there are no formal errors or mistakes in the initial specification stage, The third step, which includes modeling the system using HDL, begins. In [26], Verilog code has been applied to implement the CPS model in hardware.

To check the model functionality is correct, Matlab simulation has been used in the fourth step as a software verification of the CPS model. In this step, all the signals should be implemented in a simulation system as numerical data (test scenario). Testing the correctness of the SVM algorithm required software simulation of HDL code which is needed to use a tool that verifies the HDL code, so this paper used the Active HDL simulator tool. After hardware implemented, Wisniewski et al verify the behaviour of the implemented SVM algorithm and ensure that operate correctly by test the control signals on the I/O pins of the Field Programmable Gate Array (FPGA) device. Wisniewski et al. concluded that the results of the final stage are identical to those of the software verification step, demonstrating that the design and implementation systems match all requirements.

Also, Grobelna et al. (2017) [27], verified their approach in more than one stage. they have been applied their approach on the system for the real-life process which must be carefully and precisely constructed in order to operate flawlessly and meet the requirements of the users. To ensure the high quality of a final result, such a system requires verification methods to be applied at various stages of the system development. [27] represented a method that specifies the real-life process for distributed logic controllers using Petri-net. Petri-net can be used as modeling formalism. This methodology covers the development of distributed logic controllers, from the specification step to the final implementation stage.

The key notion of [27] was that the system is decomposed into small components (modules) that perform distributed system. Grobelna et al. (2017) [27], used model checking to formally verify the specification (before and after the decomposes process) against predefined behavioral requirements. This means that the verification step is repeated twice to guarantee that the functionality of the designed distributed logic controller functions is correct. [27] Uses the smart home system as an example to show how to design and verify a real-world process. The system should first be specified using a control interpreter Petri-net. To formally verify the primary specification, [27] used model checking techniques. In this step, an interpreted Petri-net has been translated to a rule-based logical model which is utilized to explain the core element of the Petri-net (places and transitions). The automatic generation of a verifiable model (VMOD) and synthesizable model (SMOD) are based on the algorithm outlined in [27]. Grobelna et al. (2017) [27], developed the m2vs tool to generate these models by using the rule-based logical model. The Smod is a VHDL language, and Vmod is compatible with the NuSMV model checker which compares Vmod to requirements that trans-formed to formal specifications using temporal logic, which is either LTL or CTL depending on the property type.

The net splits into several concurrent modules called SMCs during the decomposition system stage. [27] employs the ILP technique to decompose the Petri-net of a specified system. This technique can be successfully applied to most Petri-nets. Each module is a sequential automaton that can be designed separately from the rest of the prototype system. But the decomposition process suffered from the synchronization of decomposes modules problem. After the net has been decomposed and the proper synchronization has been established, the modules are written as distinct rule logical models that are automatically transformed into verified modules and then combined with these modules. Finally, [27] implemented the decomposition system in FPGA.

The behavior definition, which specifies the properties and main functionality of embedded systems, is the most significant step in embedded systems design. Errors can happen during the incoming phases, and to remove them, this process needs enormous costs. In the case of designing a dependable embedded system, if a tiny error occurred through a design stage may cause tragic effects and change the behavior of the overall system. Therefore, choosing a suitable language for modeling process is very important during the design system. The first step of system design required for gathering requirements and needing to consultations with the client. The problems that may be faced by the designer, the client may not be professional in formulating requirements. The UML language, which is simple to understand, is one of the behavior description languages. UML is also used in business areas, as well as in modeling of information flow behavioral software and embedded system design, also Petri-net, particularly Control Interpreted Petri-Net (CIPN), is used for specifying hardware behavior, but it is not easily understandable and acceptable for non-engineers. Petri-net is even used to generate code for logic controllers. Although the UML activity diagram (version 2) can be used to model hardware behavior, it is not well supported in implementation (code) creation. Grobelny et al. (2012) [28], integrated the accessibility of UML with full support for formal verification of Petri-net to improve design quality. To apply this approach, [28] supported logic controller development, starting with a specification to formal verification and finishing with synthesis. By transforming UML activity diagrams to Petri-nets and verifying with the formal approach, Grobelny et al. employed one of the verification techniques to ensure greater quality for project. To improve project quality, Grobelny et al. used formal methods such as model checking with the NuSMV tool. After specifying the properties (safety and liveness) from requirements and defining them by temporal logic (CTL or LTL), NuSMV tool has been used for checking it. Grobelny et al. state that their proposed approach was tested using control interpreted Petri nets on many examples of industrial logic controller specifications.

Another language that used for the building model is SysML which is the extension of UML developing software. SysML can model both hardware and software behaviors. SysML is made up of elements and associations which includes representations of elements, requirements, behaviors, test cases, parametrics, and the links that connect them. Khan et al. (2012) [29], used SysML to model avionics electronics and software. The approach was to extend model Based verification and validation to electronics and software through functional and structural models implemented in SysML. SysML helps to simulate hardware and software interactions. Khan et al. directly derive simulations from models and validate simulations against hardware test results. The simulation was then used to conduct test cases, and the results were compared to those obtained from identical tests run on a hardware testbed. When the simulation produced test results that were nearly identical to those of the avionics hardware, Khan et al. found that SysML model simulation at a spacecraft subsystem level could be used to mimic device functionality, and that inputs could be passed to a more capable box and card-level simulators.

As for EAST-ADL, is a language that describes the architecture of safety-critical automotive embedded system design (architectural description language). Kang et al. (2015) [30], presented a formal method for translating EAST-ADL behavior description controllers into UPPAAL models. This approach proposes an algorithm for translating a hierarchical Stateflow model to a UPPAAL model. Kang et al. presented this approach because the Stateflow was originally designed for the simulation of designs, it has limited support for formal analysis, and often suffers from incomplete coverage difficulties. As a result, it does not make the model suited for formal verification. In addition, a set of mapping rules supporting an algorithm is presented and validated to verify that the translation is correct, efficient, and applicable to real case studies. The Fault-Tolerant Fuel Control and Power Window were two automotive systems where they used in their approach. UPPAAL model checking tool has been used in [30] for real-time system modeling, validation, and verification. In UPPAAL, a subset of Timed Arithmetic Tree (TCTL) logic is used as a query language to define system requirements. As for State flow, is a graphical language similar to Statecharts in that it visualizes state machines, and the states are the primary components of it.

When Kang et al. applied their approach to the fault-tolerant fuel control system, noticed that the invalid properties do not appear through the simulation in Simulink/State flow but can be discovered directly using the UPPAAL tool which generates a counterexample for each invalid property. Kang et al. discovered the cause of invalid property after analyzing the counter-example which illustrates the errors that cause an invalid property.

As for Techniques and tools, NuSMV is a mature temporal logic model checker derived from SMV (symbolic model verification). It can verify specifications written in Computation Tree Logic (CTL) and Linear Temporal Logic (LTL). Model checkers, such as NuSMV, are frequently used for test case generation in addition to verification. Arcaini et al. (2017) [31], proposed NuSeen, a tool framework that aids designers while utilizing the NuSMV model checker. NuSeen is a set of tools that are integrated into the eclipse IDE and are designed to assist NuSMV users with their V&V activities. It focuses on making the NuSMV easier to use by using graphical components like as buttons, menus, and text highlighting, among other things. It also includes various auxiliary tools that use graphical elements to display model information (like tables, dependency graphs, etc.). NuSeen has a visualize counterexample that displays traces in a more user-friendly manner than NuSMV's basic text (or XML). The designer can better understand if the bug is in the system behavior or the temporal property by looking at the counterexample. Variables in a NuSMV specification can also be analyzed to see whether they have any dependencies. The dependency graph can be used to better understand how behaviors are related. This approach also has gaps in the editor's violation and the amount of the generated test.

NuSMV is also appropriate for large-scale models. Fritzsch et al. (2021) [32], using the NuSMV to formalize a Vehicle Control System. Several failure states had to be modeled based on the safety requirements. The required properties are then formalized and reported using CTL and LTL formulations. The model checking technique must be automated due to the system's complexity. When they describe the architecture modeling procedure, they get a large-scale model. To avoid the state explosion problem, automate and improve the model checking procedure for use on multi-core CPUs, and use Bounded Model Checking. To do this, they built an automated method for partitioning, parallelizing, and batch processing model checking. They also discussed their thoughts on synchronous vs. asynchronous timing, as well as their decision to use NuSMV. They employed Bounded Model Checking to get an acceptable runtime and solve the state explosion problem.

## 8. CONCLUSION AND FUTURE WORKS

By reviewing previous research in the field of formal verification and dependable computing, we have found that it is extremely important to implement verification and validation approach at an early stage of the system design. This approach will make the embedded digital device work efficiently, safely, and free from any interruptions, errors, or malfunctions. The previous studies which have been clarified in this paper shows that each verification approach used a different modeling language. They also differ in the application of the formal verification in terms of the type of model checking technique being used, the number of stages of the verification process throughout the system, and the level of applying the verification whether on software level, or both software and hardware levels. From this survey paper, it can help the verification researchers to propose a more robust approach than those presented in the literature review and provides a high level of verification and validation on the overall system.

TABLE I A SUMMURY OF APPROACHES FOR VERIFYING SYSTEMS PROPERTIES

| No. | Authors | Modelling | | Verification | | Criteria | | | Level | |
|---|---|---|---|---|---|---|---|---|---|---|
| | | Model Language | Property Language | Technique | Tool Support | Objective | Kind of Properties | Application | S.W | H.W |
| 1 | Askari Hemmat et al. - 2015 [19] | AutoFocus3 (AF3) | LTL-CTL | Model checking | NuSMV | automatically determining whether a design (model) satisfies desired properties | | a train control system | √ | √ |
| 2 | Kang et al. - 2017[20] | EAST-ADL-Simulink/ stateflow | LTL | Model checking /simulation | UPPAAL-SMC-Simulink/SDV | perform V&V on (non)-functional properties of an autonomous vehicle system and support for Simulink | | autonomous traffic sign recognition vehicle | √ | |
| 3 | Kumar and Kumar - 2016[21] | PROMELA Timed Automaton | LTL | Model checking | SPIN | model timed automata based hybrid systems | | TRAFFIC LIGHT CONTROLLER | √ | |
| 4 | Pakonen and Buzhinsky in [22] | MODCHK | LTL, CTL and PSL | Model checking | NuSMV/BMC | verifying the fault tolerance of I&C systems based on model checking | | instrumentation and control (I&C) | √ | √ |
| 5 | Kumar et al. in 2021 [23] | DRBD, colored Petri nets (CPN) | ------ | Model checking | DRBD, colored Petri nets (CPN) | Analyzing the (DFWCS) to locate and identify the critical aspects of reliability and safety such as nonliveness, deadlock, design errors, or faulty state | liveness | digital feedwater control system (DFWCS) of a nuclear power plant (NPP) | √ | |
| 6 | Shreya and Nanda - 2016 - [24] | Matlab/Simulink | LTL-CTL | Model checking | NuSMV | achieve Reliability and safety property of designed system and enhance the efficiency by proper verification techniques. | safety and liveness | the Mode Transition Logic (MTL) of an autopilot system. | √ | |
| 7 | Enoiu et al. – 2012-[25] | EAST-ADL | CTL-TCTL | Model checking | UPPAAL | improving the EAST-ADL language's behavioral definition, providing for formal modeling, simulation, and verification of the functional and timing requirements | safety and liveness | Brake-By-Wire (BBW) system | √ | |

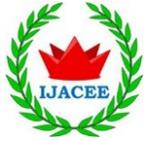

| No. | Authors | Modelling | | Verification | | Criteria | | | | |
| | | Model Language | Property Language | Technique | Tool Support | Objective | Kind of Properties | Application | Level | |
| | | | | | | | | | S.W | H.W |
| 8 | Wisniews ki et al. – 2019. [26] | Petri nets | ------ | **Model checking /simulation** | M2vs | Provide comprehensive verification for the control part of a CPS system by analyzing and verifying it in several times | | direct matrix converter (MC) with transistor commutation | √ | √ |
| 9 | Grobelna et al - 2017 - [27] | Petri nets | CTL | Model checking | NuSMV | design and verification methods of distributed logic controllers supervising real-life processes | | Smart home system | √ | √ |
| 10 | Grobelny et al. - 2012 in [28] | UML- Petri Nets | LTL-CTL | Model checking | NuSMV | transformation from UML activity diagrams into control interpreted Petri nets to Ensure higher quality of projects | | Embedded logic controller | √ | √ |
| 11 | in [29], Khan et al. - 2012 | SysML | ------ | Simulation | A Remote Engineering Unit (REU) | extends model-based V&V | | Spacecraft Avionics | √ | √ |
| 12 | Kang et al. - 2015[30] | TA—EAST-ADL | TCTL | Model checking | UPPAAL | Extend the transforming timed behavioral constraints in EAST-ADL into the analyzable UPPAAL models by including support for Stateflow | | Fault-tolerant Fuel Control (FFC) system and Power Window Controller | √ | |
| 13 | Arcaini et al. - 2017 [31] | --- | LTL-CTL | Model checking | NuSMV | Perform a framework that assists a designer during the modeling and V&V activities when using NuSMV | --- | NuSMV tool | √ | |
| 14 | Fritzsch et al. - 2021 [32] | NuSMV | LTL-CTL | Model checking | NuSMV | use the model checker NuSMV to check large-scale models with great complexity. | Safety | Vehicle Control System | √ | |

Future research works will be focusing on various topical areas, including the traditional formal verification and validation methods, such as hybrid fault-tolerant techniques and run-time verification to ensure that the behavior of the designed digital system meets its intended specification. These methods are called formal in its operation because temporal language is used in the design and analysis of system complexity. Also, we would attempt to make the designed system more reliable, safer, and operates under any potential fault by designing a fault-tolerant controller system. Finally, these designed digital systems will be verified in the implementation phase using FPGA technology.

## REFERENCES

[1] Wisniewski, Bazydło, Szcześniak, Grobelna, and Wojnakowski, "Design and Verification of Cyber-Physical Systems Specified by Petri Nets—A Case Study of a Direct Matrix Converter," *Mathematics*, vol. 7, Issue. 9, p. 812, Sep. 2019.

[2] X. Zheng, C. Julien, M. Kim, and S. Khurshid, "Perceptions on the State of the Art in Verification and Validation in Cyber-Physical Systems," *IEEE Syst. J.*, vol. 11, Issue. 4, pp. 2614–2627, Dec. 2017

[3] I. Grobelna, "Formal Verification of Control Modules in Cyber-Physical Systems," *Sensors*, vol. 20, Issue. 18, p. 5154, Sep. 2020.

[4] W. L. Oberkampf and C. J. Roy, "Verification and Validation in Scientific Computing," p. 791,2010.

[5] "IEEE Standard for System, Software, and Hardware Verification and Validation," IEEE,p.260,nov.2017.

[6] E. M. Clarke, T. A. Henzinger, H. Veith, and R. Bloem, Eds., *Handbook of Model Checking*. Cham: Springer International Publishing, 2018.

[7] P. S. Kaliappan and V. K. Kaliappan, "Deriving the behavioral properties from UML designs as LTL for model checking," in *2015 International Conference on Signal Processing, Informatics, Communication and Energy Systems (SPICES)*, Kozhikode, India, Feb. 2015, pp. 1–5.

[8] Y. Murray and D. A. Anisi, "Survey of Formal Verification Methods for Smart Contracts on Blockchain," in *2019 10th IFIP International Conference on New Technologies, Mobility and Security (NTMS)*, CANARY ISLANDS, Spain, Jun. 2019, pp. 1–6.

[9] E. A. Lee and S. A. Seshia, "Embedded Systems," p. 124,2015.

[10] "Boulanger J.-L. (ed.) - Industrial use of formal methods_ formal verification-Wiley (2012).djvu.",p.306,2012.

[11] M. Kooli, F. Kaddachi, G. D. Natale, A. Bosio, P. Benoit, and L. Torres, "Computing reliability: On the differences between software testing and software fault injection techniques," *Microprocess. Microsyst.*, vol. 50, pp. 102–112, May 2017.

[12] "Front Matter," in *Fault-Tolerant Systems*, Elsevier, 2021, pp. i–iii.

[13] R. Jhawar and V. Piuri, "Fault Tolerance and Resilience in Cloud Computing Environments," in *Cyber Security and IT Infrastructure Protection*, Elsevier, 2014, pp. 1–28.

[14] "Towards a Method for Combined Model-based Testing and Analysis:," in *Proceedings of the 2nd International Conference on Model-Driven Engineering and Software Development*, Lisbon, Portugal, 2014, pp. 609–618.

[15] D. J. Offutt, "Model-Based Testing for Embedded Systems," p. 668. Apr.2017.

[16] E. Bartocci, Y. Falcone, A. Francalanza, and G. Reger, "Introduction to Runtime Verification," in *Lectures on Runtime Verification*, vol. 10457, E. Bartocci and Y. Falcone, Eds. Cham: Springer International Publishing, 2018, pp. 1–33.

[17] I. Kurtev, J. Hooman, and M. Schuts, "Runtime Monitoring Based on Interface Specifications," in *ModelEd, TestEd, TrustEd*, vol. 10500, J.-P. Katoen, R. Langerak, and A. Rensink, Eds. Cham: Springer International Publishing, 2017, pp. 335–356.

[18] Y. Falcone, S. Krstić, G. Reger, and D. Traytel, "A Taxonomy for Classifying Runtime Verification Tools," p. 19,Sep.2018.

[19] M. H. Askari Hemmat, O. A. Mohamed, and M. Boukadoum, "Formal modeling, verification and implementation of a train control system," in *2015 27th International Conference on Microelectronics (ICM)*, Casablanca, Morocco, Dec. 2015, pp. 134–137.

[20] Eun-Young Kang, D. Mu, L. Huang, and Q. Lan, "Verification and Validation of a Cyber-Physical System in the Automotive Domain," in *2017 IEEE International Conference on Software Quality, Reliability and Security Companion (QRS-C)*, Prague, Czech Republic, Jul. 2017, pp. 326–333.

[21] N. S. Kumar and G. S. Kumar, "Modeling and verification of timed automaton based hybrid systems using spin model checker," in *2016 IEEE Annual India Conference (INDICON)*, Bangalore, India, Dec. 2016, pp. 1–8.

[22] A. Pakonen and I. Buzhinsky, "Verification of fault tolerant safety I&C systems using model checking," in *2019 IEEE International Conference on Industrial Technology (ICIT)*, Melbourne, Australia, Feb. 2019, pp. 969–974.

[23] P. Kumar, L. K. Singh, and C. Kumar, "Model Based Verification of Safety-Critical Systems," in *2021 2nd International Conference for Emerging Technology (INCET)*, Belagavi, India, May 2021, pp. 1–9. NuSMV model checker," in *2016 IEEE International Conference on Recent Trends in Electronics, Information & Communication Technology (RTEICT)*, Bangalore, India, May 2016, pp. 817–820.

[25] E. P. Enoiu, R. Marinescu, C. Seceleanu, and P. Pettersson, "ViTAL: A Verification Tool for EAST-ADL Models Using UPPAAL PORT," in *2012 IEEE 17th International Conference on Engineering of Complex Computer Systems*, Paris, Jul. 2012, pp. 328–337.

[26] Wisniewski, Bazydło, Szcześniak, Grobelna, and Wojnakowski, "Design and Verification of Cyber-Physical Systems Specified by Petri Nets—A Case Study of a Direct Matrix Converter," Mathematics, vol. 7, Issue. 9, p. 812, Sep. 2019.

[27] I. Grobelna, R. Wisniewski, M. Grobelny, and M. Wisniewska, "Design and Verification of Real-Life Processes With Application of Petri Nets," *IEEE Trans. Syst. Man Cybern. Syst.*, vol. 47, Issue. 11, pp. 2856–2869, Nov. 2017.

[28] M. Grobelny, I. Grobelna, and M. Adamski, "Hardware Behavioural Modelling, Verification and Synthesis with UML

2.x Activity Diagrams," *IFAC Proc. Vol.*, vol. 45, Issue. 7, pp. 134–139, 2012.

[29]  M. Khan, M. Sievers, and S. Standley, "Model-Based Verification and Validation of Spacecraft Avionics," presented at the Infotech@Aerospace 2012, Garden Grove, California, Jun. 2012, p.13.

[30]  E.-Y. Kang, L. Ke, M.-Z. Hua, and Y.-X. Wang, "Verifying Automotive Systems in EAST-ADL/Stateflow Using UPPAAL," in *2015 Asia-Pacific Software Engineering Conference (APSEC)*, New Delhi, Dec. 2015, pp. 143–150.

[31]  P. Arcaini, A. Gargantini, and E. Riccobene, "NuSeen: A Tool Framework for the NuSMV Model Checker," in *2017 IEEE International Conference on Software Testing, Verification and Validation (ICST)*, Tokyo, Japan, Mar. 2017, pp. 476–483.

[32]  J. Fritzsch, T. Schmid, and S. Wagner, "Experiences from Large-Scale Model Checking: Verifying a Vehicle Control System with NuSMV," in *2021 14th IEEE Conference on Software Testing, Verification and Validation (ICST)*, Porto de Galinhas, Brazil, Apr. 2021, pp. 372–382.

[33]  F. N. Kassab Bashi and S. S. Khairullah, "A Survey on Dependable Digital Systems using FPGAs: Current Methods and Challenges," International Journal of Advances in Computer and Electrical Engineering., vol. 5, issue. 12, pp. 1–8, December 2020.

[34]  S. S. Khairullah and C. R. Elks, "Self-repairing hardware architecture for safety-critical cyber-physical-systems," IET Cyber-Phys. Syst. Theory Appl., vol. 5, no. 1, pp. 92–99, March 2020.

[35]  S. S. Khairullah and A. A. Mostfa, "Reliability and Safety Modeling of a Digital Feed-Water Control System," Journal of University of Babylon for Pure and Applied Sciences JUBPAS., vol. 28, no. 1, pp. 284–304, May 2020.

[36]  S. Khairullah, "Toward Biologically-Inspired SelfHealing, Resilient Architectures for Digital Instrumentation and Control Systems and Embedded Devices," Theses Diss., pp. 1–144, December 2018.

## Authors Biography

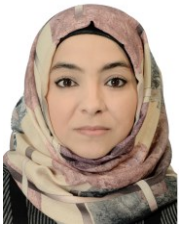

**Fayhaa Hameedi Khlaif,** received the B.S. degree in Computers and Information Engineering from the University of Nineveh, college of electronics engineering, Mosul, Iraq in 2010. Currently, she is working for M.Sc. Degree in Computer Engineering at Mosul's University College of Engineering. Her research interests include verification and validation, Formal methods, and fault-tolerant digital system design.

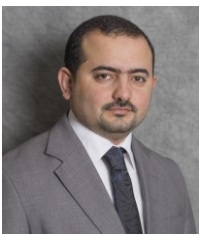

**Shawkat Sabah Khairullah,** received the B.S. and M.Sc. degrees in computer engineering from the University of Mosul, college of engineering, Mosul, Iraq in 2006 and 2011. He received the Ph.D. degree in Computer Engineering from Virginia Commonwealth University, Richmond, USA in 2018. He is a Lecturer at the University of Mosul, Iraq. Dr. Shawkat research interests include dependable digital system design, VLSI, bio-inspired self-healing hardware systems, and FPGA-based systems